\documentclass[conference]{IEEEtran}
\IEEEoverridecommandlockouts
\usepackage{cite}
\usepackage{amsmath,amssymb,amsfonts}

\usepackage{graphicx}
\usepackage{textcomp}
\usepackage{xcolor}

\usepackage{xspace}

\usepackage[utf8]{inputenc}
\usepackage[T1]{fontenc}

\usepackage[inline,shortlabels]{enumitem} 

\usepackage{algorithm}
\usepackage{algpseudocode}

\usepackage{amsmath}

\usepackage[hidelinks]{hyperref}
\usepackage{cite}

\usepackage{listings}

\newcommand{\davidechange}[1]{\textcolor{black}{#1}}

\definecolor{dark-green}{RGB}{0, 153, 76}

\algrenewcommand\textproc{}
\algblock{Input}{EndInput}
\algnotext{EndInput}
\algblock{Output}{EndOutput}
\algnotext{EndOutput}
\newcommand{\Desc}[2]{\State \makebox[1em][l]{#1}#2}
\algnewcommand\algorithmicswitch{\textbf{switch}}
\algnewcommand\algorithmiccase{\textbf{case}}
\algnewcommand\algorithmicforeach{\textbf{for each}}
\algdef{S}[FOR]{ForEach}[1]{\algorithmicforeach\ #1\ \algorithmicdo}

\usepackage{multirow}
\usepackage{booktabs}
\usepackage{tabularx}

\newcolumntype{L}[1]{>{\raggedright\let\newline\\\arraybackslash\hspace{0pt}}m{#1}}
\newcolumntype{R}[1]{>{\raggedleft\let\newline\\\arraybackslash\hspace{0pt}}m{#1}}

\usepackage{array}

\newcolumntype{C}[1]{>{\centering\arraybackslash}p{#1}}

\usepackage[most]{tcolorbox}

\usepackage{listing}
\usepackage{caption}
\usepackage{xcolor}

\let\origthelstnumber\thelstnumber
\makeatletter
\newcommand*\Suppressnumber{%
  \lst@AddToHook{OnNewLine}{%
    \let\thelstnumber\relax%
     \advance\c@lstnumber-\@ne\relax%
    }%
}

\newcommand*\Reactivatenumber{%
  \lst@AddToHook{OnNewLine}{%
   \let\thelstnumber\origthelstnumber%
   \advance\c@lstnumber\@ne\relax}%
}

\usepackage{color}
\definecolor{orange}{rgb}{1,0.647,0}
\definecolor{mygreen}{rgb}{0.30, 0.80, 0.20}

\usepackage{colortbl}
\newcommand{\hl}[1]{\cellcolor{lightgray}#1}

\newcommand{\win}[1]{\hl{#1}}
\newcommand{\draw}[1]{#1}
\newcommand{\drawall}[1]{#1}

\newtcbinputlisting{\mylisting}[2][]{%
	listing engine=listings,
	listing file={#2}, listing only,
	title={#1},
	enhanced,
	listing options={
		framerule=0.05pt,
		numbers=left,
		numberstyle=\tiny,
		basicstyle=\footnotesize\ttfamily,
		breaklines = true,
		showspaces = false,
		numbersep=2.8mm,
		numberblanklines=false
	},
	overlay={%
		\begin{tcbclipinterior}
			\fill[gray!25] (frame.south west) rectangle ([xshift=4mm]frame.north west);
		\end{tcbclipinterior}
	}
}

\usepackage[htt]{hyphenat}

\makeatletter
\def\BState{\State\hskip-\ALG@thistlm}
\makeatother
\makeatletter
\newcommand{\rqitem}[2][]{%
	\ifblank{#1}{%
		\item #2%
	}{%
		\item[#1] #2%
	}%
	\protected@edef\@currentlabelname{#2 (\theenumi)}

}
\makeatother

\newcommand{\approach}{{\sc Except}\xspace}
\newcommand{\approachbold}{{Except}\xspace}
\newcommand{\numFautsException}{{33}\xspace}
\newcommand{\numCodeFaults}{{43}\xspace}

\newcommand{\toolUrl}{\url{https://gitlab.com/qrs2021/except}\xspace}

\begin{document}
\title{\textit{Exception-Driven Fault Localization for Automated Program Repair}\\
%
}

\author{\IEEEauthorblockN{Davide Ginelli, Oliviero Riganelli, Daniela Micucci, Leonardo Mariani}
\IEEEauthorblockA{
\textit{University of Milano - Bicocca}, Milan, Italy \\
\{davide.ginelli, oliviero.riganelli, daniela.micucci, leonardo.mariani\}@unimib.it}
}

\maketitle


\begin{abstract}
Automated Program Repair (APR) techniques typically exploit spectrum-based fault localization (SBFL) to identify the program locations that should be patched, making the effectiveness of APR techniques dependent on the effectiveness of fault localization. Indeed, results show that SBFL often does not localize faults accurately, hindering the effectiveness of APR.

In this paper, we propose \approach, a technique that addresses the localization problem by focusing on the \emph{semantics} of failures rather than on the correlation between the executed statements and the failed tests, as SBFL does. We focus on failures due to exceptions and we exploit their type and source to localize and guess the faults.
Experiments with \numCodeFaults exception-raising faults from the Defects4J benchmark show that \approach can perform better than Ochiai and ssFix.
\end{abstract}

\begin{IEEEkeywords}
automatic program repair, fault localization, SBFL, exceptions
\end{IEEEkeywords}

\section{Introduction} \label{sec:introduction}

Automated Program Repair (APR) techniques have been extensively studied as approaches that can assist developers during the debugging and bug fixing tasks~\cite{Gazzola:Repair:TSE:2017,Goues:APR:CACM:2019}. Although often limited to the simplest bugs, APR has already demonstrated its effectiveness with large-scale industrial systems~\cite{Marginean:SapFix:ICSE:2019}.



APR techniques offer a range of strategies to repair code. For example, \emph{generate-and-validate} techniques (e.g., GenProg~\cite{le2012genprog} and AE~\cite{weimer2013leveraging}) discover plausible fixes by exploring a space of potential patches, while \emph{semantics-driven} techniques (e.g., SemFix~\cite{nguyen2013semfix} and NOPOL~\cite{xuan2017nopol}) synthesize fixes from a representation of the repair problem directly derived from the faulty application.


APR techniques share the challenge of identifying the \emph{fix locus}, that is, the program location(s) that should be modified to produce a fix. Indeed, it is \emph{hard or even impossible} to repair a fault without selecting a good location for the fix~\cite{Liu:ICST:APRLocalization:2019}.

The larger the program size, the more difficult it is to identify the correct location(s) in which to apply the fix. Focusing on the wrong locations may waste significant computational resources, since each location can be modified in many different ways in the attempt of obtaining a fix~\cite{Long:SearchSpace:ICS:2016}. Even worse, choosing the wrong location multiple times can dramatically impact performance and even the feasibility of the repair process. 

APR techniques address the problem of identifying the fix locus using Spectrum-Based Fault Localization (SBFL)~\cite{Wong:SurveyLocalization:TSE:2016}, which can heuristically compute a suspiciousness score associated with each program location: the higher the score, the higher the probability that the program location is faulty. The suspiciousness scores of the statements is derived from the information about the number of passing and failing test cases that execute them. Intuitively, the more a statement is executed by failing executions only, the more suspicious it is.

The suspiciousness scores associated with the statements determine a ranking that is used by APR techniques to identify the targets of the repair process. For example, jGenProg~\cite{MARTINEZ201965}, uses the ranking generated by the Ochiai SBFL technique~\cite{Abreu:Ochiai:TAICPART:2007} to select statements as modification points with a probability that depends on their suspiciousness, 
while NOPOL~\cite{xuan2017nopol} follows the ranking generated by Ochiai to analyze each statement in the ranking one after the other. Unfortunately, experimental evidence shows that SBFL techniques are often unable to rank faulty statements at top positions~\cite{Liu:ICST:APRLocalization:2019,Assiri:Localization:SQJ:2016,Pearson:EvaluationFL:ICSE:2017}.


In this paper, we propose to address the localization problem by primarily exploiting the \emph{semantics of the failures} rather than the correlation between the executed statements and the failed tests, that has been proven to produce inaccurate results. 
To this end, we focus on \emph{failures caused by exceptions}, which represent a large portion of the failures that can be observed\davidechange{~\cite{Sawadpong2012, ginelli2020comprehensive, repairnator}}. 


\emph{Exceptions carry extensive information about the 
occurred failures}, such as the location that raised the exception, which can represent a good starting point for fault localization, and the type of the exception, which provides useful semantic information about the possible nature of the problem. For instance, a failure caused by a Java {\small\texttt{ArrayIndexOutOfBoundsException}} suggests the statement that raised the exception as a possible location for the fix, but the statement where the array has been initialized, and the statements that assigned a value to the variables used to access a location of the array are good locations as well. The type of the exception can thus be used to guide the analysis in the selection of the suspicious locations according to the semantics of the failure.

Since locations are identified based on a guessed cause of the failure (e.g., the value of an index might be wrong), a suspicious location can be enriched with information about the program elements that are likely responsible for the failure (e.g., a variable in an index expression) and the likely fault (e.g., wrong variable used), which can in turn be exploited to identify the change that should be operated to correct the program (e.g., replace the variable). 


Although some approaches addressed the localization and repair of failures caused by exceptions, {\small \texttt{NullPointerExceptions}} in particular~\cite{Duriez:NullPointerPatch:SANER:2017,Sinha:ExceptionsRepair:ISSTA:2009}, they cannot generate ranked lists of statements that can be used by APR techniques, neither they considered enriching the localization with debugging information that can be useful to developers. The localization used in ssFix~\cite{Xin:ssFix} can exploit stack traces to improve rankings, but this is not sufficient to correctly localize several faults, as reported in our evaluation.


This paper presents \approach, an exception-driven fault localization technique that can be used to support APR techniques. 
\approach analyzes failures caused by exceptions and localizes faults by following the semantics of exceptions and their causes.
In addition to producing an \emph{ordered list of statements} that are reported as high-priority elements on top of ranked lists returned by SBFL techniques, \approach enriches the identified items with information about the \emph{individual expressions} that are likely faulty (e.g., a specific variable in an expression) and the \emph{guessed faults} (e.g., wrong variable used), which can be used by APR techniques and developers to determine how to patch programs (e.g., replacing the variable). 

In our empirical evaluation, \approach outperformed both Ochiai~\cite{Abreu:Ochiai:TAICPART:2007} and the localization used in ssFix~\cite{Xin:ssFix} for \numCodeFaults exception-raising fault locations selected from the Defects4J benchmark~\cite{Just:2014:DDE:2610384.2628055}. Further, \approach could correctly guess faults for multiple exception types. 



In a nutshell, the main contributions of this paper are:
\begin{itemize}[leftmargin=*]
\item \approach, a fault localization technique that considers the semantics of exceptions to localize faults for APR, enriching the localization with information about the \emph{expressions} 
likely responsible for the failures and the \emph{guessed faults},
\item Empirical evidence of the effectiveness of the proposed strategy in comparison to Ochiai and ssFix.
\end{itemize}

This paper is organized as follows. Sections~\ref{sec:approach} presents \approach. Section~\ref{sec:exceptions} describes the supported exceptions. Section~\ref{sec:evaluation} presents the empirical evaluation. Section~\ref{sec:related} discusses the related approaches. Section~\ref{sec:conclusion} provides final remarks.
\section{Except}
\label{sec:approach}

\begin{figure*}[ht]
	\centering
	\includegraphics[width=0.99\textwidth]{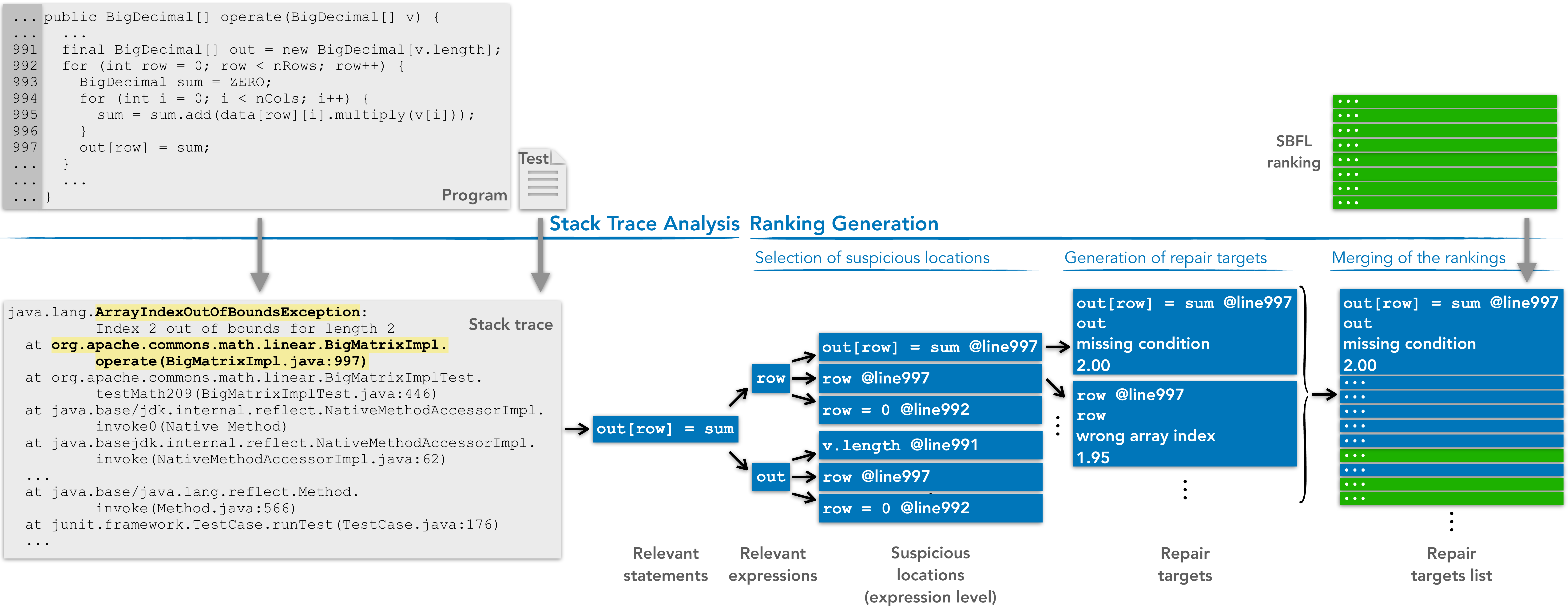}
	\vspace{-6pt}
	\caption{\approachbold applied to the program in Listing~\ref{lst:index-out-of-bounds}.}
	\label{fig:strategy}
\end{figure*}

\begin{algorithm}[ht]
	\caption{Description of \approach.}\label{alg:approach}
	\footnotesize
	\begin{flushleft}
	\begin{algorithmic}[1]
		\Procedure{\approach(p, t, rankingSBFL)}{}
			\Input
				\Desc{p}{\hspace*{4.7em} The faulty program}
				\Desc{t}{\hspace*{4.7em} A test case that fails raising an exception}
				\Desc{rankingSBFL}{\hspace*{4.7em} The list of suspicious statements identified with a \hspace*{9.1em}SBFL technique}
			\EndInput
			
			\Output
				\Desc{repairTargetsList: a list of repair target or null}
			\EndOutput 
			\item[]
			\State \color{dark-green}// Step 1: Stack Trace Analysis\color{black}
			\State stackTrace = getExceptionStackTrace(p, t); \label{alg1:step1B}
			\State exceptionType = getExceptionType(stackTrace);
			\State relevantStatementsList = getRelevantStatementsList(stackTrace); \label{alg1:step1E} \label{alg1:output-step1}
			\item[]
					
			\State  \color{dark-green}// Step 2: Ranking Generation\color{black}
			
			\State repairTargetsList = $\varnothing$;  \label{alg1:step2B}
			\State relevantStmsAnalyzed = 0;
			\item[]
			
			\State  \color{dark-green}// Sub-step 2.1: Selection of suspicious locations\color{black}
			\ForEach{relevantStatement $\in$ relevantStatementsList} \label{alg1:step2.1B}
				\If{relevantStmsAnalyzed < \hspace*{\fill} \color{blue}exceptionType.maxRelevantStatementsToConsider\color{black}}\label{alg1:step2.1:NRS}
				
					\State relevantStmsAnalyzed = relevantStmsAnalyzed + 1;  			
					\State suspiciousLocations = selectSuspiciousLocations(p, \hspace*{\fill} relevantStatement, exceptionType); \label{alg1:step2.1E}
			\State  \color{dark-green}// Sub-step 2.2: Generation of repair targets\color{black}
					\ForEach{suspLoc $\in$ suspiciousLocations} \label{alg1:step2.2B} 
		
						\State susp = computeSuspValue();
						\State repairTarget = \color{blue} generateRepairTarget \color{black} (suspLoc, \hspace*{\fill} exceptionType, susp);
						
						\State repairTargetsList.add(repairTarget);\label{alg1:step2.2E}	
					\EndFor
				\EndIf
			\EndFor
			\item[]
			\State  \color{dark-green}// Sub-step 2.3: Merging of the rankings\color{black}
			\State repairTargestList.addTargetsFrom(rankingSBFL); \label{alg1:step2E}	
			\item[]
					
			\State \Return repairTargetsList;
			
	\EndProcedure
\end{algorithmic}
\end{flushleft}
\end{algorithm}

As exemplified in Figure~\ref{fig:strategy}, \approach returns a ranked list of \emph{repair targets} starting from three inputs: a faulty program \emph{p}, a test \emph{t} that fails with an uncaught exception, and a list of suspicious statements \emph{rankingSBFL} identified with an SBFL technique. A repair target reports the following information: 
\begin{itemize}[leftmargin=*] 
\item a program \emph{location}, which is a likely faulty program statement,
\item an \emph{expression} in the location, which represents a specific program element likely responsible of the fault,  
\item \emph{guessed faults}, which associate the expression with specific guessed faults that may affect the expression,  
\item a \emph{suspiciousness value}, which is a positive number that can be used to rank repair targets from the most likely to the least likely to be relevant for fixing the fault.
\end{itemize}

For example, a repair target may refer to an array variable (expression) in a statement with an array access (program location) as a likely faulty element, while suggesting that the array name is wrong (guessed fault), with a given suspiciousness value. Or, it may identify the variable used as index of the array (expression) as the faulty element, suggesting that the value of the variable is wrong (guessed fault), with a given suspiciousness value.

\approach adds the repair targets that derive from the knowledge of the exception raised by the failing test as high priority items in the initial ranked list produced by a SBFL technique. Repair targets and SBFL locations are merged together to obtain a comprehensive ranked list that can benefit from the joint contributions of two complemental approaches. Since the information about the faulty expression and the guessed fault derives from the knowledge of the exceptions, they are available only for the high priority targets added by \approach, and are not available for the elements in the initial ranked list produced by SBFL.



\approach works in two main steps: 
\begin{itemize}[leftmargin=*]
\item \emph{Stack Trace Analysis}, which analyzes the stack trace of the exception to identify the \emph{type of the exception} and the \emph{relevant statements} that occur in the stack trace; 
\item \emph{Ranking Generation}, which identifies the \emph{relevant expressions} that may have caused the exception, traces them back to the \emph{suspicious locations} that might have affected the values of the relevant expressions, and creates a \emph{ranked list of repair targets} from the suspicious locations and the input SBFL ranking.
\end{itemize}

Figure~\ref{fig:strategy} shows the elements that are incrementally identified by \approach to finally generate the ranking, when applied to the faulty program in Listing~\ref{lst:index-out-of-bounds}, which throws an  {\small \texttt{Array\-Index\-Out\-Of\-Bounds\-Exception}}.
\autoref{alg:approach} and \autoref{alg:find-susp-locs} detail how \approach works with pseudocode. The blue constant and the blue functions 
depend on the type of the exception and are described in details in Section~\ref{sec:exceptions} for the supported exceptions.

\subsection{Stack Trace Analysis}

Stack trace analysis (lines \ref{alg1:step1B}-\ref{alg1:step1E} of \autoref{alg:approach}) extracts two key data from the stack trace associated with an exception: the \emph{exception type} and the \emph{relevant statements}. 

The exception type is explicitly reported in the stack trace and can be trivially retrieved.





\approach identifies as relevant statements that might have contributed to the exception every program statement explicitly reported in the exception stack trace, that is, every location that appears in the context of the statement that raised the exception. 
Since the fault is assumed to be in the program, \approach discards the statements that do not refer to the program under analysis, but rather refer to external libraries, JDK classes, test frameworks (e.g., JUnit~\cite{jUnit} or Mockito~\cite{mockito}), and test cases. 
For each unfiltered statement, \approach creates a \emph{relevant statement} which includes
the \emph{line number} of the statement, the \emph{Java class} to which it belongs to, the \emph{method} that executes it, and the \emph{file name} containing the statement. This is done by function {\small \texttt{getRelevantStatementsList}} invoked at line \ref{alg1:output-step1} of \autoref{alg:approach}.

\begin{algorithm}
	\caption{Description of selectSuspiciousLocations.}\label{alg:find-susp-locs}
	\footnotesize
	\begin{flushleft}
	\begin{algorithmic}[1]
		\Procedure{selectSuspiciousLocations(p, stp, et)}{}
		\Input
		\Desc{p}{\hspace*{1em} The faulty program}
		\Desc{stp}{\hspace*{1em} A Stack Trace Poi}
		\Desc{et}{\hspace*{1em} The type of the exception} 
		\item[]
		\EndInput
		
		\Output
		\Desc{suspiciousLocations: a set of suspicious locations}
		\EndOutput 
		\item[]
		
		\State relevantExpressions = \color{blue}selectRelevantExpressions\color{black}(p, stp, et);  \label{alg2:step1}
		\item[]
		\State suspiciousLocations = $\varnothing$;
		
		\ForEach{re $\in$ relevantExpressions}
		
		\State suspiciousLocationsForRe = \color{blue}findSuspiciousLocations\color{black}(p, re, et); \label{alg2:step2}
		\State suspiciousLocations.add(suspiciousLocationsForRe);
		
		
		\EndFor
		\item[]
		
		\State \Return suspiciousLocations;
		
		\EndProcedure
	\end{algorithmic}
	\end{flushleft}
\end{algorithm}


The list of relevant statements is ordered according to their position in the stack trace, starting from the one closest to the statement that generates the exception under analysis.

\subsection{Ranking Generation}


The generation of the ranking (lines \ref{alg1:step2B}-\ref{alg1:step2E} in \autoref{alg:approach}) implies \emph{1)} analyzing the relevant statements to identify the suspicious program locations that might have caused the exception, 
 \emph{2)} generating the repair targets,
and  \emph{3)} merging the identified repair targets with the initial SBFL ranking.

\textbf{Selection of Suspicious Locations} (lines \ref{alg1:step2.1B}-\ref{alg1:step2.1E} in \autoref{alg:approach} and \autoref{alg:find-susp-locs}). The number of the analyzed relevant statements is bound to prevent that too many repair targets can be generated (line \ref{alg1:step2.1:NRS} in \autoref{alg:approach}). In fact, adding many targets to the initial ranking generated by SBFL may hinder the effectiveness of APR techniques, since they would have to consider too many highly suspicious program locations. On the contrary, \approach aims to add a \emph{small and focused} set of high-priority repair targets that may help in directing repair algorithms on the right statements for the right reason.

In practice, the number of relevant statements to be considered might be different based on the exception type. 
The bound could be small (e.g., 1) for  some exceptions. For instance, in the case of {\small \texttt{ArrayIndexOfOutBoundsException}}, the statement that raises the exception includes the array variable and the index value that causes the exception and the analysis can be effectively driven by their values. The bound could be higher for other exceptions, such as the {\small \texttt{IllegalArgumentException}}, since the raised exception may strongly depend on the execution context, and considering multiple points derived from the stack trace of the exception might be beneficial (e.g., also considering the calling method).

The selection of the suspicious locations from a relevant statement is described in the {\small \texttt{selectSuspiciousLocations}} function presented in \autoref{alg:find-susp-locs}. The selection is driven by two key logical steps: the \emph{selection of the relevant expressions} (line \ref{alg2:step1} in \autoref{alg:find-susp-locs}) and the \emph{identification of the suspicious locations} (line \ref{alg2:step2} in \autoref{alg:find-susp-locs}). 

When a relevant statement is analyzed, \approach first narrows down the analysis to the specific expressions that might be responsible for the exception, ignoring the rest of the statement. For example, 
if a statement raises an {\small \texttt{ArrayIndexOutOfBoundsException}}, \approach would select the expression used to access the array and the expression that identifies the array as relevant expressions for the analysis, as shown in Figure~\ref{fig:strategy}. 
Since this step of the analysis depends on the semantics of the exception, it is described in details for each supported exception in Section~\ref{sec:exceptions}.

Each expression relevant to the exception 
is used to identify the statements that might have caused the exceptional situation that finally resulted in the failure. For instance, 
if the relevant expressions are the array name and the array index, \approach would select the code that defines the array and the code that defines the index as suspicious locations. This is done with a local data-flow analysis that depends on the relevant expression and the type of exception. The specific analysis for the supported exceptions is described in Section~\ref{sec:exceptions}.

\textbf{Generation of Repair Targets} (lines \ref{alg1:step2.2B}-\ref{alg1:step2.2E} of \autoref{alg:approach}).
Every suspicious location is turned into a repair target by adding the guessed faults and a suspiciousness value. The guessed faults are annotations that specify why the expression could be faulty. For example, if the exception is {\small \texttt{ArrayIndexOutOfBoundsException}} and the selected expression defines the value of the size used to initialize the array, a guessed fault may assume the initial size of the array is wrong. Developers and APR techniques can exploit this annotation to change the program accordingly. Since the annotation depends on the type of the exception, we describe how faults are guessed for each exception type in Section~\ref{sec:exceptions}.

The suspiciousness value assigned to the repair targets considers the fact that these targets must have higher priority compared to the top ranked items in the input SBFL ranking. Since the maximum suspiciousness in the input SBFL ranking is 1, \approach assigns suspicious values that start from 2 to the identified repair targets. 
Since repair targets are generated by following the order of occurrence of the relevant statements, which are ordered based on their distance from the statement that raises the exception, \approach prioritizes the repair targets accordingly, decreasing the suspiciousness value by 0.05 every time a target is added to the ranking. This gap is chosen to accomodate up to 20 high-priority repair targets before potentially reaching the elements in the input ranking. 

\textbf{Merging of the rankings} (line \ref{alg1:step2E} of \autoref{alg:approach}).
The last step requires merging the identified repair targets with the input ranked list. This is performed with two simple steps. First, the locations that are present both in the SBFL ranking and in the repair targets are removed, keeping only the location with the highest suspiciousness score. The additional information produced by \approach (the expression and the guessed fault) are also preserved. Second, all the items are ordered according to their suspiciousness. 

\section{Supported Exceptions} \label{sec:exceptions}

To demonstrate \approach, we defined the analysis for four types of exceptions: {\small \texttt{ArrayIndexOfOutBoundsException}}, {\small \texttt{StringIndex\-Out\-Of\-BoundsException}}, {\small \texttt{NullPointer\-Exception}}, and {\small \texttt{IllegalArgumentException}}. 
We focused on some of the most popular types of exceptions based on faults contained in public benchmarks, such as Defect4J~\cite{Just:2014:DDE:2610384.2628055}, Bears~\cite{Madeiral2019}, and Repairnator~\cite{repairnator}. Similar analyses can be added to support additional exceptions.

\approach is equipped with simple and fast analyses based on data flow and bounded in scope to identify the likely fault locations. In particular, the analysis to determine the suspicious locations is bounded to the method that includes the relevant expression. Note that multiple relevant expressions in multiple methods can be selected, thus the analysis is not generally limited to the method that raises the exception. Moreover, the scope of the analysis always includes the definition of class variables, which may initialize variables with wrong values. Bounding the analysis is useful to generate a limited number of repair targets and complete the analysis quickly (in our experiments every  case could be processed in few seconds). 

In the following, we discuss and exemplify the elements that depend on the exceptions for the four supported exception types: {\small \texttt{maxRelevantStatementsToConsider}} specifies the number of relevant statements to consider, {\small \texttt{selectRelevantExpressions}} describes how the relevant expressions are determined, {\small \texttt{findSuspiciousLocations}} describes the analysis that selects the suspicious code locations, and {\small \texttt{generateRepairTarget}} indicates the guessed faults that are associated with the suspicious locations. The guessed fault consists of a label with known semantics (e.g., ``wrong variable name'') that is included in the repair target and that can be exploited by developers or APR techniques. 

\subsection{ArrayIndexOfOutBoundsException}
\noindent \texttt{\textbf{maxRelevantStatementsToConsider}}. The analysis only considers the statement that raises the exception. 

\smallskip
\noindent \texttt{\textbf{selectRelevantExpressions}}. \approach looks for instances of the expression \texttt{refArray[exprIndex]} in the relevant statement, to select the occurrences of \texttt{refArray} and \texttt{exprIndex} as relevant expressions that might be wrong and thus cause the exception. In fact, the access to the array might fail because the wrong array is used or the wrong array location is selected.  

\smallskip
\noindent \textbf{\texttt{findSuspiciousLocations} and \texttt{generateRepairTarget}}. When \texttt{refArray} is considered, \approach runs a recursive backward bounded data-flow analysis to identify the locations in which \texttt{refArray} is allocated. If the array initialization statement uses other variables, the analysis process is iterated to determine the locations that assign a value to these variables. Also the location itself, where \texttt{refArray} is used, is returned as a suspicious location. 
%

When \texttt{exprIndex} is considered, \approach runs a backward bounded data-flow analysis to identify the locations that define the variables that occur in \texttt{exprIndex}. Also the location itself where \texttt{exprIndex} is used is returned as a suspicious location. 
Table~\ref{tab:refArray} lists the identified locations and the corresponding guessed faults.
%


\begin{table} 
\caption{Analysis of {\footnotesize \texttt{ArrayIndexOutOfBoundsException}}.}
	\label{tab:refArray}
	\footnotesize
	\renewcommand{\arraystretch}{0.4} 
\begin{tabular}{p{4.5cm}p{3.5cm}}
	\toprule
		\textbf{Suspicious locations} & \textbf{Guessed faults} \\
	\midrule
	\multirow{2}{*}{\textit{statement with} \texttt{refArray}} & array variable is wrong\\ 
	 & missing conditional statement\\ 
	\midrule
	\textit{allocation}\ \textit{of} \texttt{refArray} & wrong array initialization\\
	\midrule
	\textit{definitions of the variables that determine the size of} \texttt{refArray} & \multirow{2}{*}{wrong variables values} \\
	\midrule
	\textit{statement with}\ \texttt{exprIndex} & \texttt{exprIndex} is wrong\\ 
	\midrule
	\textit{definitions of variables used in} \texttt{exprIndex} &  {wrong variables values}\\ 
	\bottomrule
\end{tabular}
\end{table}


%


\begin{lstlisting}[caption={Example of \footnotesize{\texttt{ArrayIndexOutOfBoundsException}}.},label={lst:index-out-of-bounds},language=Java,escapechar=|,numbers=left,xleftmargin=1.8em,framexleftmargin=1.6em,framexrightmargin=-0.1em]
public BigDecimal[] operate(BigDecimal[] v) { 
	...     
	final BigDecimal[] out = new BigDecimal[v.length]; |\label{line:array-initialization}|
	for (int row = 0; row < nRows; row++) { |\label{line:row-assignment}|
		BigDecimal sum = ZERO;
		for (int i = 0; i < nCols; i++) {
			sum = sum.add(data[row][i].multiply(v[i]));
		}
		out[row] = sum; |\label{line:array-assignment}|
	}
} /*** Source: Math 98 - Defects4J ***/
\end{lstlisting}

\smallskip
\noindent \textbf{Example}. \autoref{lst:index-out-of-bounds} shows an excerpt of the Math 98 fault in Defects4J. The statement at line~\ref{line:array-assignment} is the relevant statement, since it generates the {\small \texttt{ArrayIndexOutOfBoundsException}}. The variables {\small \texttt{out}} and {\small \texttt{row}} are selected as \emph{relevant expressions}, which in turn generate 6 \emph{suspicious locations} with the corresponding guessed fault: {\small \texttt{out}} might be the wrong array variable used at line~\ref{line:array-assignment}; {\small \texttt{row}} might be the wrong index used at line~\ref{line:array-assignment}; there might be a missing condition (e.g., it is necessary to add an {\small \texttt{if-statement}} that influences the access to array {\small \texttt{out}}); {\small \texttt{row}} might be assigned with the wrong value at line~\ref{line:row-assignment}; the initialization of {\small \texttt{out}} at line~\ref{line:array-initialization} might be wrong; the expression {\small \texttt{v.length}} used at line~\ref{line:array-initialization} might be wrong. In this case, the correct fix consists in modifying the initialization of the array at line~\ref{line:array-initialization} by replacing the expression {\small \texttt{v.length}} with {\small \texttt{nRows}}, which is one of the guessed faults.

\subsection{StringIndexOutOfBoundsException}
\noindent \textbf{\texttt{maxRelevantStatementsToConsider}}. 
The analysis only considers the statement that raises the exception. 

\smallskip
\noindent \textbf{\texttt{selectRelevantExpressions}}. 
\approach looks for instances of the expression \texttt{stringVar.op(...exprIndex...)} in the relevant statement, where \texttt{stringVar} is a \texttt{String} variable, \texttt{op} is a method that can return a {\small \texttt{StringIndexOutOfBoundsException}}, such as {\small \texttt{charAt(int index)}}, and \texttt{exprIndex} is an Integer expression for accessing the string at a specific position. 

\approach selects the occurrences of \texttt{stringVar}, in case the wrong string is used, and \texttt{exprIndex}, in case the wrong index is used, as relevant expressions.

\smallskip 
\noindent \textbf{\texttt{findSuspiciousLocations} and \texttt{generateRepairTarget}}. When either \texttt{stringVar} or \texttt{exprIndex} are considered, \approach runs a backward data-flow analysis to identify the locations in which the variables included in these expressions are defined. 
 Also the location where these variables are used is returned as a suspicious location. 
 Table~\ref{tab:StringExcep} lists the identified locations and the guessed faults.
 
 \begin{table}
\caption{Analysis of {\footnotesize \texttt{StringIndexOutOfBoundsException}}.}
\label{tab:StringExcep}
	\footnotesize
		\renewcommand{\arraystretch}{0.4} 
\begin{tabular}{p{4.5cm}p{3.5cm}}
	\toprule
	\textbf{Suspicious locations} & \textbf{Guessed faults} \\
	\midrule
	\multirow{2}{*}{\textit{statement with} \texttt{stringVar}} & \texttt{String} variable is wrong\\ 
	  & missing conditional statement\\ 
	\midrule
	\multirow{2}{*}{\textit{definition of} \texttt{stringVar}} & wrong value\\
		  & missing conditional statement\\ 
	\midrule
	\textit{statement with} \texttt{exprIndex} & \texttt{exprIndex} is wrong\\ 
	\midrule
	\multirow{2}{*}{\textit{definitions of} \texttt{vars} \textit{used in} \texttt{exprIndex}} & wrong variables values\\
		  & missing conditional statement\\ 
	\bottomrule
\end{tabular}
\end{table}


\begin{lstlisting}[caption={Example of {\footnotesize \texttt{StringIndexOutOfBoundsException}}.},label={lst:string-index-out-of-bounds},language=Java,escapechar=|,numbers=left,xleftmargin=1.8em,framexleftmargin=1.6em,framexrightmargin=-0.1em]
public static String abbreviate(String str, int lower, int upper, ...) {
		...
		if (upper == -1 || upper > str.length())
		    upper = str.length();
		if (upper < lower)
    		upper = lower;
		...
		if (index == -1) {
    		result.append(str.substring(0, upper)); |\label{line:substring}|
		    ...
		}
} /*** Source: Lang 45 - Defects4J ***/
\end{lstlisting}

\smallskip
\noindent \textbf{Example}. \autoref{lst:string-index-out-of-bounds} shows an excerpt of the Lang 45 fault in Defects4J. The statement at line~\ref{line:substring} is the relevant statement, since it generates the {\small \texttt{StringIndexOutOfBoundsException}}. The analysis selects the expressions {\small \texttt{str}}, {\small \texttt{0}} and {\small \texttt{upper}} as \emph{relevant expressions}, which in turn generate 8 \emph{suspicious locations} with the corresponding guessed fault. For sake of space, we do not list all of them, but they include the actual fix, which consists of a missing condition.

\subsection{NullPointerException}
\noindent \texttt{\textbf{maxRelevantStatementsToConsider}}. The analysis considers both the statement that raises the exception and the calling method to address the case of \texttt{null} values erroneously passed as parameters. 

\smallskip
\noindent \texttt{\textbf{selectRelevantExpressions}}. \approach looks for instances of the expression \texttt{obj.op()} in the first relevant statement, where \texttt{obj} is a non-primitive variable. 
 \approach selects the occurrences of \texttt{obj} as relevant expressions that might be wrong and thus cause the exception. 
Finally, \approach selects the non-primitive parameters used in the method call in the second relevant statement to account for \texttt{null} values generated by the caller. 


\smallskip 
\noindent \textbf{\texttt{findSuspiciousLocations} and \texttt{generateRepairTarget}}.  When \texttt{obj} is considered, \approach runs a backward bounded data-flow analysis to identify the locations where \texttt{obj} is defined. When the caller is analyzed, if the \texttt{null} variable is defined through the method call, the calling site is also identified as a suspicious location. 
 The location itself where this variable is used is also returned as a suspicious location. 
 Table~\ref{tab:NullExcep} lists the identified locations and the corresponding guessed faults.

 \begin{table}
\caption{Analysis of {\small \texttt{NullPointerException}}.}
\label{tab:NullExcep}
	\footnotesize
		\renewcommand{\arraystretch}{0.4} 
\begin{tabular}{p{4.5cm}p{3.5cm}}
	\toprule
		\textbf{Suspicious locations} & \textbf{Guessed faults} \\	
		\midrule
	\multirow{2}{*}{\textit{statement with} \texttt{obj}} & variable is wrong\\ 
	  & missing conditional statement\\ 
	\midrule
	\multirow{2}{*}{\textit{definition of} \texttt{obj}} & wrong value\\
		  & missing conditional statement\\ 
	\midrule
	\textit{calling}\ \textit{site} & wrong variables\\ 
	\bottomrule
\end{tabular}
\end{table}

%

\begin{lstlisting}[caption={Example of {\small \texttt{NullPointerException}}.},label={lst:null-pointer},language=Java,escapechar=|,numbers=left,xleftmargin=1.8em,framexleftmargin=1.6em,framexrightmargin=-0.1em]
	public class XYPlot {
 		...
 		public Range getDataRange(ValueAxis axis) {
  			...
  			XYItemRenderer r = getRendererForDataset(d); |\label{line:r-assignment}|
  			...
  			Collection c = r.getAnnotations(); |\label{line:npe}|
  			...
 		}
} /*** Source: Chart 4 - Defects4J ***/
\end{lstlisting}

\smallskip
\noindent \textbf{Example}. \autoref{lst:null-pointer} shows an excerpt of the Chart 4 fault included in Defects4J. The statement at line~\ref{line:npe}, since it generates the {\small \texttt{Null\-Pointer\-Exception}}, is a relevant statement. The analysis selects the expression \texttt{r} as \emph{relevant expression}, which in turn generates 3 \emph{suspicious locations} with the corresponding guessed fault: the variable {\small \texttt{r}} at line~\ref{line:npe} might be wrong; the entire statement at line~\ref{line:npe} might have to be accessed only within a conditional statement; and the method {\small \texttt{getRendererForDataset()}} used to assign a value to the variable {\small \texttt{r}} at line~\ref{line:r-assignment} might be wrong. The actual fix consists of adding the conditional statement.





\subsection{IllegalArgumentException}
\noindent \texttt{\textbf{maxRelevantStatementsToConsider}}. \approach considers both the statement that raises the exception and the caller statement of the method that raises the exception as relevant statements for the analysis. 

\smallskip
\noindent \texttt{\textbf{selectRelevantExpressions}}. \approach looks for one or more instances of the expression \texttt{op(...exprPar...)} in the relevant statements. The expression represents any method call with at least one parameter. If the invocation is found in the first relevant statement, the second relevant statement is skipped. Otherwise, (e.g., the first relevant statement throws the exception instead of invoking a method), the second relevant statement is also searched for the same pattern. 

 \approach selects the occurrences of \texttt{exprPar} as relevant expressions that might be wrong, thus causing the exception. 
 
\smallskip 
\noindent \textbf{\texttt{findSuspiciousLocations} and \texttt{generateRepairTarget}}. 
When \texttt{exprPar} is considered, \approach runs a backward bounded data-flow analysis to identify the locations where the variables used in \texttt{exprPar} are defined. 
Table~\ref{tab:IllegalExcep} lists the identified locations and the guessed faults.

 \begin{table}
\caption{Analysis of {\small \texttt{IllegalArgumentException}}.}
\label{tab:IllegalExcep}
	\footnotesize
		\renewcommand{\arraystretch}{0.4} 
\begin{tabular}{p{4.5cm}p{3.5cm}}
	\toprule
		\textbf{Suspicious locations} & \textbf{Guessed faults} \\
	\midrule
	\multirow{2}{*}{\texttt{exprPar} \textit{used as parameter}} & wrong parameter\\
		  & wrong method invoked\\ 
	\midrule
	\textit{definition of variables in} \texttt{exprPar}& wrong value\\
	\bottomrule
\end{tabular}
\end{table}



\begin{lstlisting}[caption={Example of {\small \texttt{IllegalArgumentException}}.},label={lst:illegal-argument},language=Java,escapechar=|,numbers=left,xleftmargin=1.8em,framexleftmargin=1.6em,framexrightmargin=-0.1em]
	public class TimeSeries {
  		...
  		public Object clone() {
    		Object clone = createCopy(0, getItemCount() - 1); |\label{line:create-copy-call}|
    		return clone;
  		}
  
  	public TimeSeries createCopy(int start, int end) throws ... {
    	if (start < 0)
       		throw new IllegalArgumentException("");
    	if (end < start)
       		throw new IllegalArgumentException(""); |\label{line:iae}|
    	...
  	}
} /*** Source: Chart 17 - Defects4J ***/
\end{lstlisting}

\smallskip
\noindent \textbf{Example}. \autoref{lst:illegal-argument} shows an excerpt of the Chart 17 fault in Defect4J. The statement at line~\ref{line:iae} raises an {\small \texttt{IllegalArgumentException}} if the value of the second parameter of method {\small \texttt{createCopy}} is less than the first one. The corresponding relevant locations are the statements at lines~\ref{line:iae} and~\ref{line:create-copy-call}. The former statement does not contribute to the analysis since it does not include the actual invocation, while the latter does.
The suspicious locations derived from the latter statement are four: the invocation of method {\small \texttt{createCopy(int, int)}}; the integer {\small \texttt{0}}; the expression {\small \texttt{getItemCount() - 1}}; and the call to {\small \texttt{getItemCount()}}. The fix requires changing the expression {\small \texttt{getItemCount() - 1}} at line~\ref{line:create-copy-call}.




\section{Empirical Evaluation}
\label{sec:evaluation}

The empirical evaluation aims to answer three research questions. 

\begin{enumerate}[start=1, label={\bfseries RQ\arabic*}, ref={RQ\arabic*}, before=\bfseries, align=left, wide=0pt, leftmargin=0em]

	\rqitem{What is the fault localization effectiveness of \approachbold ?} \label{rq1} \textnormal{This research question investigates the fault-localization effectiveness of \approach for different types of exceptions in comparison to state of the art solutions.}
	

	\rqitem{How does \approachbold affect the capability of modifying the faulty statements of APR techniques that use SBFL?} \label{rq2} \textnormal{This research question investigates how using the ranking produced by \approach affects the capability of program repair techniques to modify the right statement every time the generation of a patched program is attempted.}

	\rqitem{What is the accuracy of the guessed fault?} \label{rq3} \textnormal{This research question investigates how accurately \approach can guess the fault that generated an exception.}

\end{enumerate}

\subsection{Empirical Setup}

In the evaluation, we used faults from the Defects4J benchmark v1.5~\cite{Just:2014:DDE:2610384.2628055}, a dataset of real-world bugs and fixes from 5 open-source Java projects (JFreeChart, Closure Compiler, Commons Math, Joda-Time, and Commons Lang) commonly used in program repair studies~\cite{Xin:ssFix,Martinez:APR:2017,Kui:TBar:2019}.

We initially selected every bug that fails with a test that raises an uncaught exception. The four most frequent exception types raised by these bugs match with the classes of exceptions supported by \approach: {\small \texttt{NullPointerException}} (17 faults), {\small \texttt{IllegalArgumentException}} (15 faults), {\small \texttt{Array\-Index\-Out\-Of\-Bounds\-Exception}} (10 faults), {\small \texttt{String\-Index\-Out\-Of\-Bounds\-Exception}} (7 faults). Out of these 49 faults, we discarded the faults that require changing multiple non-contiguous code locations without having a same number of failing tests because they could not be properly addressed by localization techniques. For instance, we keep faults that require changing 2 locations if there are 2 failing tests that can be used to localize these changes, but we discard the same fault if only one failing test case is available. We ended up with \numFautsException faults and \numCodeFaults fault locations to be localized.
\autoref{tab:ochiai-vs-except} column \emph{Bug ID} reports the id of the selected faults, organized by exception type. In case a fault requires changing multiple locations, the table reports multiple rows for the same Bug ID.


\smallskip

To answer RQ1 and RQ2, we considered two competing approaches: the Ochiai SBFL technique~\cite{Abreu:Ochiai:TAICPART:2007}, which is the most used fault localization technique in program repair~\cite{Liu:ICST:APRLocalization:2019}, and the localization strategy used in ssFix~\cite{Xin:ssFix} (hereafter referred as \emph{ssFix}), which exploits the entries in the stack trace as top items of the ranking. Differently from \approach, ssFix does not analyze the target program to extract the relevant expressions, thus missing to select the suspicious program locations that are not in the stack trace.

To precisely measure the improvement that \approach and ssFix can introduce over an existing ranking, we used the ranking generated by Ochiai as input ranking for both \approach and ssFix. We used the test cases available with the programs to compute the ranking with Ochiai.
Our tool, which uses Spoon~\cite{pawlak:hal-01169705} for static program analysis, and experimental material are available at \toolUrl.


\begin{table}[!ht]
\caption{Effectiveness results.} \vspace{-0.2cm}
\label{tab:ochiai-vs-except}
	\scriptsize
		\renewcommand{\arraystretch}{0.4} 
		\setlength{\tabcolsep}{4pt}
		\begin{center}
	\begin{tabular}{lcccccc}

	\toprule
	\multirow{2}[3]{*}{\textbf{Bug ID}} &
	\multicolumn{3}{c}{\bfseries Position} & \multicolumn{2}{c}{\bfseries Probability (\%)} & \textbf{Additional}  \\
	\cmidrule(lr){2-4} \cmidrule(lr){5-6} & Ochiai & ssFix & \approach & Ochiai & \approach & \textbf{Info}\\
	
	\bottomrule
	&&&&&&\\
	
	\multicolumn{7}{l}{\texttt{\textbf{ArrayIndexOutOfBoundsException}} (\textbf{AIOOBE})}\\
	\cmidrule{1-4}
	
	\multirow{2}{*}{Lang 12} & \draw{6.50} & 7.50 & \draw{6.50} & \win{6.61} & 5.57 & None \\
	& \draw{6.50} & 7.50 & \draw{6.50} & \win{6.61} & 5.57 & None \\ \cmidrule(r){1-4} \cmidrule(l){5-7} 
	
	Lang 61  & \win{20.00} & 21.00 & 21.00 & \win{2.57} & 2.31 & None \\  \cmidrule(r){1-4} \cmidrule(l){5-7} 
	Math 3  & 9.50 & 10.00 & \win{2.00} & 5.58 & \win{17.65} & Only Target \\  \cmidrule(r){1-4} \cmidrule(l){5-7} 
	\multirow{2}{*}{Math 98}  & 10.00 & 10.50 & \win{2.00} & 3.19 & \win{10.43} & Both  \\  
	 & 10.00 & 10.50 & \win{2.00} & 3.19 & \win{10.4}3 & Both \\  \cmidrule(r){1-4} \cmidrule(l){5-7} 
	Mockito 34 & \draw{52.00} & \draw{52.00} & 53.00 & \win{0.30} & 0.26 & Both \\  
	
	\bottomrule
	&&&&&&\\
	\multicolumn{7}{l}{\texttt{\textbf{StringIndexOutOfBoundsException}}  (\textbf{SIOOBE})}\\
	\cmidrule{1-4}
	
	Lang 6  & 118.50 & 2.00 & \win{1.00} & 0.54 & \win{7.15} & Both \\  \cmidrule(r){1-4} \cmidrule(l){5-7} 
	Lang 44  & \draw{90.50} & \draw{90.50} & 94.00 & \win{0.81 }& 0.65 & None \\  \cmidrule(r){1-4} \cmidrule(l){5-7} 
	Lang 45  & \draw{11.50} & \draw{11.50} & 13.00 & \win{1.22} & 1.04 & Only Guess \\  \cmidrule(r){1-4} \cmidrule(l){5-7} 
	Lang 51  & - & - & - & 0.00 & 0.00 & - \\  \cmidrule(r){1-4} \cmidrule(l){5-7}
	Lang 59 & 3.50 & \draw{1.00} & \draw{1.00} & 14.61 & \win{14.85} & Both \\  \cmidrule(r){1-4} \cmidrule(l){5-7}
	Math 101  & 9.00 & \draw{1.00} & \draw{1.00} & 1.89 & \win{8.68} & Both \\  
		
	\bottomrule
	&&&&&&\\
	\multicolumn{7}{l}{\texttt{\textbf{NullPointerException}} (\textbf{NPE})}\\
	\cmidrule{1-4}
	
	Chart 4  & 53.00 & 2.50 & \win{1.00} & 0.12 & \win{0.45} & Both \\  \cmidrule(r){1-4} \cmidrule(l){5-7} 
	\multirow{4}{*}{Chart 14}  & 17.50 & 19.50 & \win{1.00} & 0.66 & \win{2.18}  & Both \\
	& 17.50 & 19.50 & \win{1.00} & 0.66 & \win{2.18} & Both \\
	& 17.50 & 19.50 & \win{1.00} & 0.66 & \win{2.18} & Both \\
	& 17.50 & 19.50 & \win{1.00} & 0.66 & \win{2.18} & Both \\  \cmidrule(r){1-4} \cmidrule(l){5-7} 
	Closure 2 & 6.00 & 5.00 & \win{1.00} & 0.33 & \win{2.63} & Both \\  \cmidrule(r){1-4} \cmidrule(l){5-7} 
	\multirow{2}{*}{Lang 20} &  23.50 & 23.50 & \win{1.00} & 2.43 & \win{9.86} & Both  \\
	 & 8.00 & 11.00 & \win{1.00} & 3.47 & \win{12.17} & Both \\  \cmidrule(r){1-4} \cmidrule(l){5-7} 
	Lang 33  & 4.50 & \draw{1.00} & \draw{1.00} & 7.96 & \win{13.74} & Both \\  \cmidrule(r){1-4} \cmidrule(l){5-7} 
	Lang 39  & 27.50 & 1.50 & \win{1.00} & 1.92 & \win{5.14} & Both \\  \cmidrule(r){1-4} \cmidrule(l){5-7} 
	\multirow{2}{*}{Lang 47}  & 82.50 & \draw{1.00} & \draw{1.00} & 0.86 & \win{6.78}  & Both \\
    	& 86.50 & \draw{1.00} & \draw{1.00} & 0.74 & \win{6.78} & Both \\  \cmidrule(r){1-4} \cmidrule(l){5-7} 
	Lang 57  & 3.50 & \draw{1.00} &\draw{1.00} & 13.46 & \win{18.05} & Both  \\  \cmidrule(r){1-4} \cmidrule(l){5-7} 
	\multirow{2}{*}{Math 4}  & 4.50 & \win{2.00 }& 3.00 & 0.82 & \win{32.48}  & Only Target \\
	 & 4.50 & \win{1.50} & 8.50 & 0.82 & \win{3.51} & None \\  \cmidrule(r){1-4} \cmidrule(l){5-7} 
	Mockito 18  & \drawall{644.00} & \drawall{644.00} & \drawall{644.00} & \draw{0.11} & \draw{0.11} & None \\  \cmidrule(r){1-4} \cmidrule(l){5-7} 
	\multirow{3}{*}{Mockito 35}  & \drawall{1.00} & \drawall{1.00} & \drawall{1.00} & \draw{0.83} & \draw{0.83} & None \\  
	 & \drawall{33.00} & \drawall{33.00} & \drawall{33.00} & \draw{0.31} & \draw{0.31} & None \\  
	& \drawall{5.00} & \drawall{5.00} & \drawall{5.00} & \draw{0.69} & \draw{0.69} & None \\  \cmidrule(r){1-4} \cmidrule(l){5-7} 
	Mockito 36  & 4.00 & \draw{1.00 }& \draw{1.00} & 0.93 & \win{5.79} & Both \\  \cmidrule(r){1-4} \cmidrule(l){5-7} 
	Mockito 38  & 1.50 & 1.50 & \win{1.00} & 0.86 & \win{5.87} & Both \\  \cmidrule(r){1-4} \cmidrule(l){5-7} 
	Math 70 & \win{1.00} & 2.00 & 3.00 & \win{14.04} & 9.03 & None \\   
	
	\bottomrule
	&&&&&&\\
	\multicolumn{7}{l}{\texttt{\textbf{IllegalArgumentException}} (\textbf{IAE})}\\
	\cmidrule{1-4}
	
	Chart 9 &  \win{13.50} & 14.50 & 16.50  &\win{ 1.50 } & 1.21 & None \\  \cmidrule(r){1-4} \cmidrule(l){5-7} 
	Chart 13 &  52.50 & 3.50 & \win{1.00} & 0.74 & \win{1.46} & Both \\  \cmidrule(r){1-4} \cmidrule(l){5-7} 
	Chart 17 &  5.00 & 1.50 & \win{1.00} & 2.38 & \win{7.59} & Only Target \\ \cmidrule(r){1-4} \cmidrule(l){5-7} 
	Chart 24  & \draw{5.50} & 6.00 & \draw{5.50} & \draw{7.36} & \draw{7.36} & None \\  \cmidrule(r){1-4} \cmidrule(l){5-7} 
	Closure 19  & - & - & - & 0.00 & 0.00 & - \\  \cmidrule(r){1-4} \cmidrule(l){5-7} 
	Lang 5  & \drawall{38.00} & \drawall{38.00} & \drawall{38.00} & \draw{1.22} & \draw{1.22} & None \\  \cmidrule(r){1-4} \cmidrule(l){5-7} 
	Lang 54  & \drawall{82.50} & \drawall{82.50} & \drawall{82.50} & \draw{0.86} & \draw{0.86} & None \\  \cmidrule(r){1-4} \cmidrule(l){5-7} 
	Time 27  & \draw{2,636.00} & 2,638.00 & \draw{2,636.00} & \win{0.0037} & 0.0036 & None \\
	\bottomrule
\end{tabular}
\end{center}
\end{table}


\subsection{What is the fault localization effectiveness of \approach?}

In this research question, we compare the fault localization effectiveness of Ochiai, ssFix, and \approach. We considered the location modified by developers as the correct fault location to be identified. If new code is added, the location next to the added code is considered as the correct one. 
The metric that we used to compare the approaches is the \emph{position} of the faulty statement in the ranking. If the faulty statement has the same suspiciousness of other statements, we consider its average position, as done in other studies. 



\autoref{tab:ochiai-vs-except} column \emph{Position} reports the ranking of the faulty statement for Ochiai, ssFix, and \approach. When a technique outperforms another, we highlight the cell with light grey. 


In the case of {\small \texttt{ArrayIndexOutOfBoundsException}}, \approach ranked the faulty statement better than competing approaches three times. Ochiai performed better than others once (Lang 61), while ssFix was never the best approach. Interestingly, while Ochiai introduced a marginal improvement for Lang 61, \approach significantly improved the ranking, moving to the second position faulty statements that were below the ninth position in the initial ranking. 

In the case of {\small \texttt{StringIndexOutOfBoundsException}}, \approach ranked the faulty statement better than competing approaches once, while there was not a single winning approach in the rest of the cases. In two cases, Ochiai and ssFix performed better than \approach, although with a marginal improvement on the raking. On the contrary, \approach significantly improved the Ochiai ranking in two cases. For this class of exceptions, ssFix and \approach performed similarly. In one case, (Lang 51), none of the approaches could locate the correct statement.


In the case of {\small \texttt{NullPointerException}}, \approach performed significantly better than both Ochiai and ssFix both in the number of cases (10 cases) and the magnitude of the improvement. Vice versa, Ochiai and ssFix performed better than the other approaches in 1 and 2 cases, respectively, with marginal improvement in the ranking compared to \approach.

Finally, in the case of {\small \texttt{IllegalArgumentException}}, \approach obtained the best result in 2 cases, while Ochiai obtained the best result in 1 case, and ssFix never obtained the best result. Again, \approach obtained a significative relative improvement over Ochiai, while the improvement of Ochiai over \approach was marginal. Also in this case, one fault related to the addition of new code was impossible to localize.


Overall, it is noticeable how \approach managed to rank well a number of faults compared to Ochiai and ssFix, with ssFix performing better than Ochiai but worse than \approach. 


\subsection{How does \approach affect the capability of modifying the faulty statements of APR techniques that use SBFL?}

To answer RQ2, we consider how APR techniques use the rankings returned by SBFL techniques. There are two main possible models: \emph{probabilistic} and \emph{one-by-one}. 

In the probabilistic model, APR techniques assign to each statement a probability to be selected for mutation that is proportional to its suspiciousness and the suspiciousness of the rest of the statements in the ranking:

\begin{equation}
	\label{eq:prob-jgenprog}
	prob(s) = \frac{susp(s)}{\sum\limits_{s_i \in ranking} susp(s_i)}
\end{equation}

\noindent where $s$ is a statement in the ranking and $ranking$ is the considered ranking. The higher the probability of selecting the faulty statement is, the more likely the APR technique shall modify the right code location to fix the fault.  Examples of APR techniques that use this probabilistic schema to iteratively identify the statement to be modified are jGenProg~\cite{MARTINEZ201965}, jMutRepair~\cite{martinez:hal-01321615}, DeepRepair~\cite{8668043}, and Cardumen~\cite{10.1007/978-3-319-99241-9_3}.

In the one-by-one model, APR techniques consider the statements in the same order they occur in the ranking from the most suspicious to the least suspicious. The lower the position value of the faulty statement in the ranking is, the sooner the APR technique will modify the right code location to fix the fault. Example of APR techniques that use this systematic schema to iteratively identify the statement to be modified are NOPOL~\cite{xuan2017nopol}, AE~\cite{weimer2013leveraging}, and jKali~\cite{martinez:hal-01321615}. 

\begin{figure}[ht]
	\includegraphics[width=0.5\textwidth]{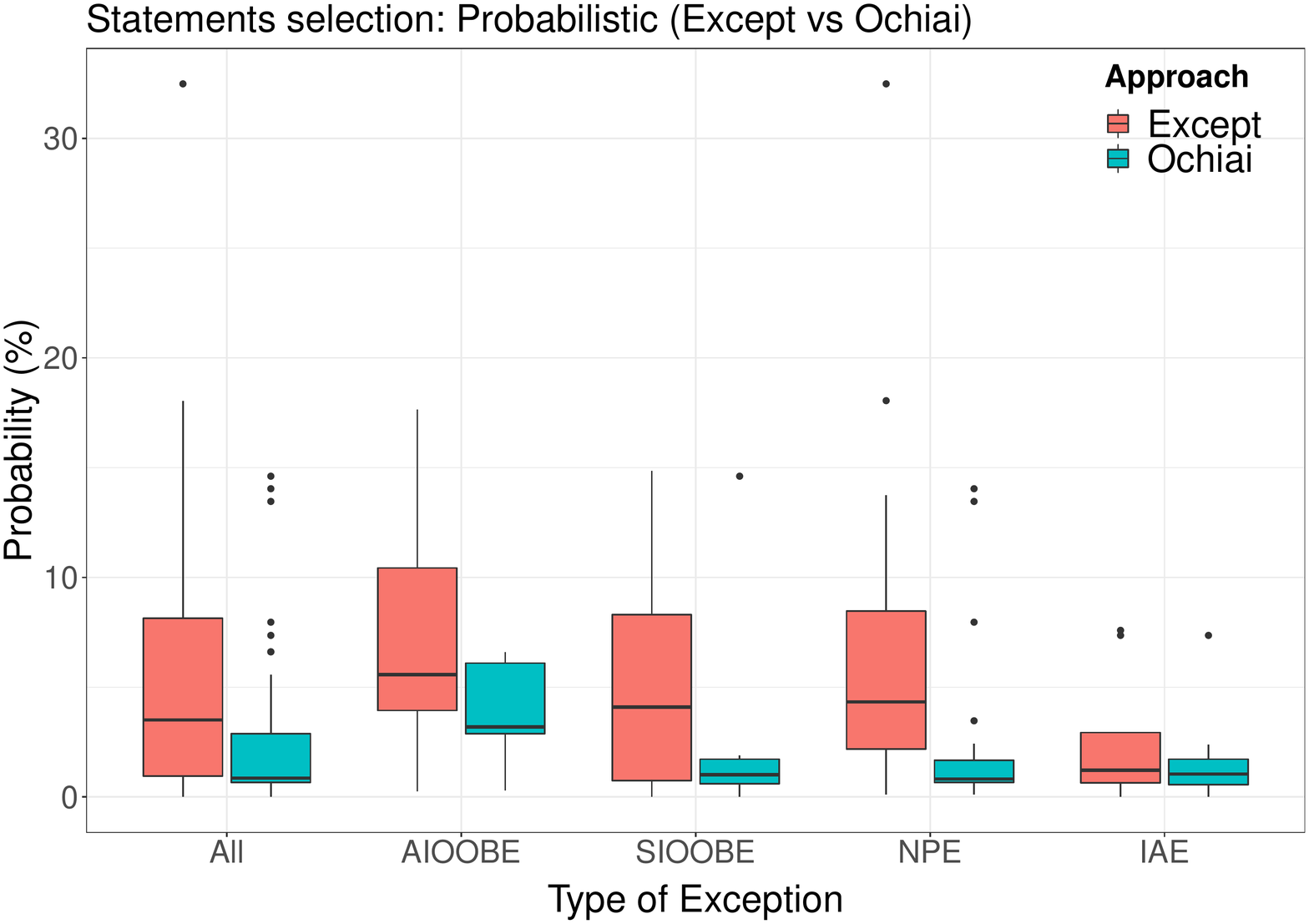}
	\caption{Comparison between \approachbold and Ochiai according to the \emph{probabilistic} usage of the ranking.}
	\label{fig:prob-boxplots-prob}
\end{figure}

\begin{figure}[ht]
	\includegraphics[width=0.5\textwidth]{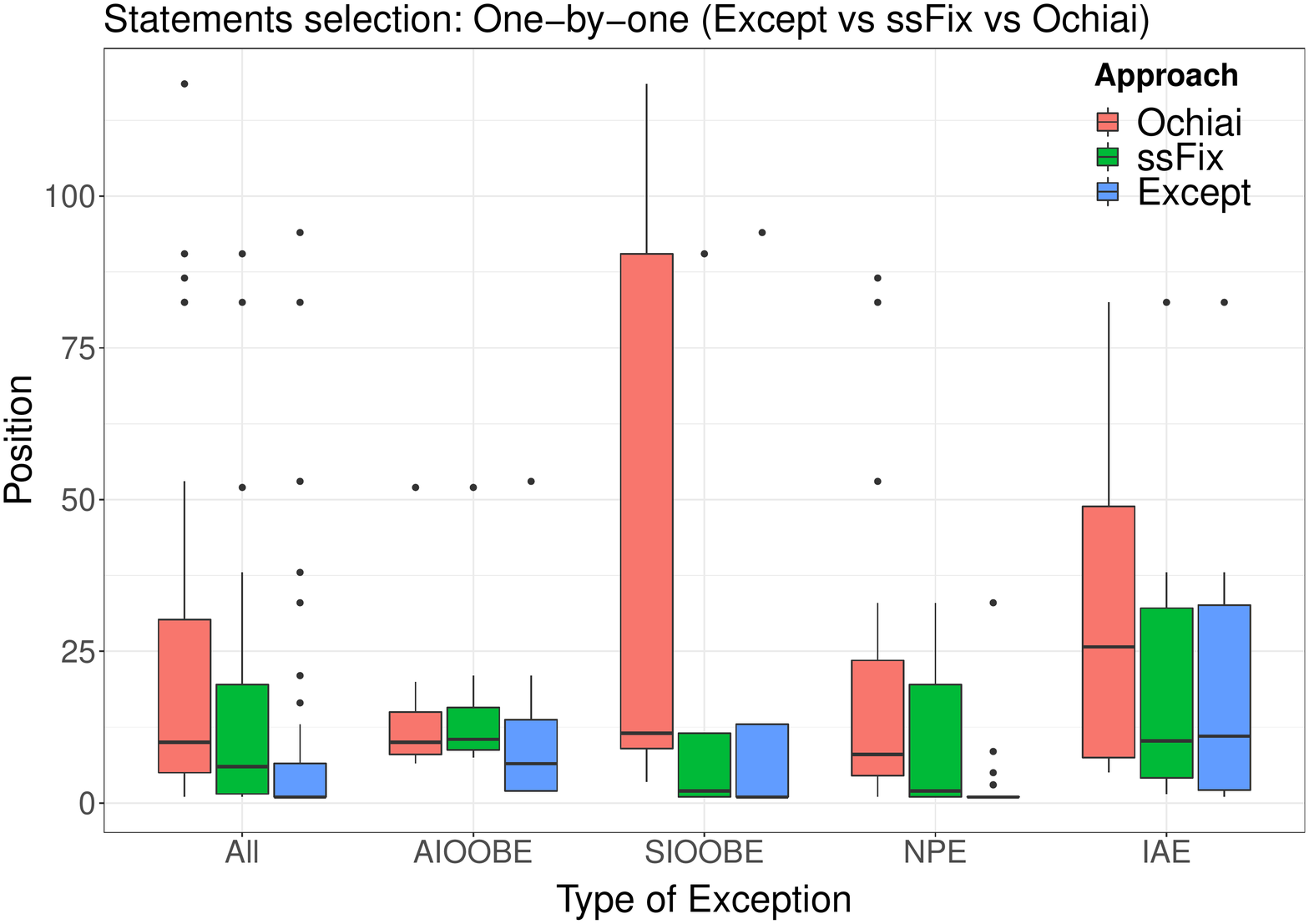}
	\caption{Comparison between \approachbold, ssFix, and Ochiai according to the \emph{one-by-one} usage of the ranking.}
	\label{fig:prob-boxplots-1by1}
\end{figure}

To assess the impact of the rankings generated by \approach, ssFix, and Ochiai in the context of APR, we computed the metrics $prob(s)$ and $position(s)$ for all the faulty statements $s$ considered in our study. Since ssFix does not assign a probability to the statements moved at the top of the ranking, we could only assess the rankings produced by ssFix with the \emph{position} metric. The analytical values of these metrics are reported in the columns \emph{Position} and \emph{Probability} of Table~\ref{tab:ochiai-vs-except}. 
%

Figure~\ref{fig:prob-boxplots-prob} visually shows the results obtained by considering the probabilistic selection schema, distinguishing per exception type and overall across all the exceptions. Probability values increase significantly with \approach. In fact, the median, third quartile, and maximum obtained with \approach are 3.51\%, 8.14\%, and 32.48\% respectively, while the median, third quartile, and maximum obtained with Ochiai are 0.86\%, 2.88\%, and 14.61\% respectively. The significant difference ($\alpha=0.05$) between \approach and Ochiai with all exceptions has been confirmed with a Wilcoxon rank sum test. 

The better performance of \approach is also observable at the level of the individual exception types, although with variable strength. In particular, while differences are remarkable for {\small \texttt{Array\-Index\-Out\-Of\-Bounds\-Exception}}, {\small \texttt{String\-Index\-Out\-Of\-Bounds\-Exception}} and 
\linebreak
{\small \texttt{Null\-Pointer\-Exception}}, the difference is smaller for {\small \texttt{Illegal\-Argument\-Exception}}. 

Figure~\ref{fig:prob-boxplots-1by1} visually shows the results obtained considering the one-by-one selection schema, distinguishing per exception type and overall across all the exceptions. The relative performance of the three approaches was confirmed, with \approach performing better than ssFix which performed better than Ochiai.  The significant difference ($\alpha=0.05$) between \approach and both Ochiai and ssFix with all exceptions has been again confirmed with a Wilcoxon rank sum test. 

At the level of the individual exceptions, we can observe that \approach performed better than ssFix and Ochiai for both {\small \texttt{Array\-Index\-Out\-Of\-Bounds\-Exception}} and {\small \texttt{Null\-Pointer\-Exception}}, while it performed comparably to ssFix for {\small \texttt{String\-Index\-Out\-Of\-Bounds\-Exception}} and 
\linebreak
{\small \texttt{Illegal\-Argument\-Exception}}.

In a nutshell, the outcome of this research question suggests that faults revealed by tests that generate uncaught exceptions should be addressed by APR techniques using the ranking produced by \approach, regardless of the strategy used (probabilistic or one-by-one). 

\subsection{What is the accuracy of the guessed fault?}

This research question evaluates the capability of \approach to guess the fault that should be repaired. We assess the accuracy of the repair target distinguishing four cases: both the selected expression and the guessed faults associated with the faulty statement are correct (label \texttt{Both}), the faulty expression is correctly selected but the guessed fault is wrong (label \texttt{Only} \texttt{Target}), the fault is correctly guessed but the wrong expression is selected (label \texttt{Only} \texttt{Guess}), and both the selected expression and the guessed fault are wrong (label \texttt{None}). \autoref{tab:ochiai-vs-except} column \emph{Additional Info} shows the results.

%

\approach both identified the expression to be changed and the fault to be fixed in 22 out of the 41 cases with a localization (54\%). In three cases the faulty expression was identified but without correctly guessing the fault. In one case, \approach guessed the right fault, but associated it with the wrong statement (\approach guessed a missing condition, although not in the right place). In total, \approach correctly enriched the localization with additional information in 26 out of 41 cases (63\%). This result indicates that the additional information generated by \approach frequently represents a meaningful suggestion about the fault. 




\subsection{Threats to validity}
\label{sec:threats}




\davidechange{A threat to validity is about the limited set of exceptions supported by \approach and the generalizability of the results. To mitigate this threat, we selected the exceptions by considering common types of problems reported in popular Java benchmarks, such as Defect4J~\cite{Just:2014:DDE:2610384.2628055}, Bears~\cite{Madeiral2019}, and Repairnator~\cite{repairnator}, that contain real-world faults from different projects raising different types of exceptions.}

Another concern is about the correctness of the implementations that we used in the experiments. We relied on GZoltar~\cite{6494960}, a widely used tool, for the implementation of Ochiai. We instead used our own implementation of ssFix~\cite{Xin:ssFix}. 
To mitigate any implementation threat, we made our artefacts publicly available.


\section{Related Work}
\label{sec:related}

The classes of approaches closely related to \approach are SBFL, slice-based, and stack trace-based techniques. 




\davidechange{\emph{Spectrum-Based Fault Localization (SBFL)} is probably the most widely used fault localization method ~\cite{Wong:SurveyLocalization:TSE:2016}. It computes a suspiciousness score for the statements executed by a program, assigning an higher value to the statements mostly executed by failing test cases.}
\davidechange{Fault localization approaches based on SBFL are often unable to rank the faulty statements at the very top positions of the rankings~\cite{Assiri:Localization:SQJ:2016,Liu:ICST:APRLocalization:2019,Pearson:EvaluationFL:ICSE:2017} and they do not provide information about why a given location might be considered faulty, complicating their usage~\cite{parnin2011automated}.} 



\approach addresses these limitations for failures caused by uncaught exceptions. In particular, \approach exploits the semantics of the failures to improve the ranking and enrich locations with information about the likely faulty expressions and the guessed faults.


\emph{Slice-based techniques} use data and control-flow dependencies to focus the localization process on the slice of program locations that may affect the output of a test case~\cite{Lei:COMPSAC:2012,Mao:JSS:2014,Weiser:slice:1986}. The basic idea is that if a test case fails due to the incorrect value of a variable, the fault should be searched, either statically~\cite{Gyimothy:ESEC:99,Weiser:slice:1986} or dinamically~\cite{Agrawal:PLDI:1990,Mao:JSS:2014}, in the slice of statements that may affect that variable.

Slicing techniques tend to select many locations, not significantly reducing the search space in practical situations. 
\approach exploits static data-flow oriented slicing to identify the suspicious locations from the relevant expressions, but heavily bounds the scope of the analysis to select a small number of statements likely causing the exception. 

\emph{Stack trace-based fault localization} methods analyze the exception stack trace in order to improve fault localization. For example, Wong et al. \cite{Wong:StackTrace:2014} analyze the stack traces that occur in bug reports to increase the suspiciousness of files. Similarly, ssFix \cite{Xin:ssFix} prioritizes the statements that occur in the exception stack trace. However, as shown in our experimentation, using the statements that explicitly occur in stack traces is not always sufficient to effectively localize faults. CrashLocator~\cite{Wu:CrashLocator:2014} can localize faults from stack traces using static analysis, based on a large collection of stack traces collected during failures. \approach also exploits stack traces, but it can be conveniently applied to a single failing execution.

\section{Conclusion}
\label{sec:conclusion}

Fault localization is a key component of APR techniques, in fact a fault that is not localized cannot be repaired~\cite{Liu:ICST:APRLocalization:2019}. 




\davidechange{This paper presents \approach, a technique that exploits the semantics of the exception and the related stack trace, to identify a small set of highly suspicious statements that are considered with high priority by APR techniques. The results obtained with our experiments show the effectiveness of \approach, also in comparison with the other approaches.}


Future work concerns with both experimenting \approach with APR techniques and studying solutions that can exploit the semantics of failures also when no exception is raised.

\bibliographystyle{IEEEtran}
\bibliography{IEEEabrv,main}

\end{document}